\documentclass{sigchi}

\CopyrightYear{2022}
\setcopyright{rightsretained}

\conferenceinfo{arXiv,}{April 2022}

\usepackage{balance}       
\usepackage{graphics}      
\usepackage[T1]{fontenc}   
\usepackage{txfonts}
\usepackage{mathptmx}
\usepackage[pdflang={en-US},pdftex]{hyperref}
\usepackage{color}
\usepackage{booktabs}
\usepackage{textcomp}

\usepackage[printonlyused, nolist]{acronym} 
\usepackage{color}
\usepackage[caption=false]{subfig}   

\usepackage{microtype}        
\usepackage{ccicons}          

\usepackage{todonotes}

\def\plaintitle{EMMI: Empathic Human-Machine Interaction for Establishing Trust in Automated Driving}

\def\emptyauthor{}
\def\plainkeywords{trust in automation; trust detection; gaze detection; automated driving; HMI; UX}

\makeatletter
\def\url@leostyle{%
  \@ifundefined{selectfont}{
    \def\UrlFont{\sf}
  }{
    \def\UrlFont{\small\bf\ttfamily}
  }}
\makeatother
\def\pprw{8.5in}
\def\pprh{11in}

\definecolor{linkColor}{RGB}{6,125,233}
\hypersetup{%
  pdftitle={\plaintitle},
  pdfauthor={\emptyauthor},
  pdfkeywords={\plainkeywords},
  pdfdisplaydoctitle=true, 
  bookmarksnumbered,
  pdfstartview={FitH},
  colorlinks,
  citecolor=black,
  filecolor=black,
  linkcolor=black,
  urlcolor=linkColor,
  breaklinks=true,
  hypertexnames=false
}

\begin{document}

\title{\plaintitle}

\numberofauthors{1}
\author{
\alignauthor{Tobias Oetermann$^{1}$, Pia Dautzenberg$^{1}$, Gudrun Voß$^{1}$, Christopher Brockmeier$^{1}$, Stefan Ladwig$^{1}$, Patrick Gebhard$^{2}$, Tanja Schneeberger$^{2}$, Markus Funk$^{3}$, Raymond Brueckner$^{3}$, Felix Schäfer$^{4}$, Norbert Helff$^{5}$, Andreas Gomer$^{6}$, Lutz Eckstein$^{1}$\\
\affaddr{Institute for Automotive Engineering (ika) RWTH Aachen University$^{1}$, German Research Center for Artificial Intelligence (DFKI)$^{2}$, Cerence GmbH$^{3}$, CanControls GmbH$^{4}$, Charamel GmbH$^{5}$,  Saint-Gobain Research Germany$^{6}$}\\
}
}

\maketitle

\begin{abstract}
Highly automated vehicles represent one of the most crucial development efforts in the automotive industry. In addition to the use of research vehicles, production vehicles for the general public are realistic in the near future. However, to fully exploit the benefits of these systems, it is fundamental that users have an appropriate level of trust in automation. Recent studies indicate that more research is needed in this area. Furthermore, beyond the management of user trust, the system should also convey a perceptible added value to realize  not only trust, but also acceptance and thus use. The EMMI project pushes both, the management of user trust while conveying added value to the user. Therefore, an advanced socio-emotional user model for estimating user trust and various user-centered HMI systems with unique UX are being developed. Together, the systems are employed to induce changes in the users' trust in automated vehicles.
\end{abstract}

\ccsdesc[500]{Human-centered computing~Human computer interaction (HCI)}
\ccsdesc[100]{Human-centered computing~HCI theory, concepts and models}

\keywords{\plainkeywords}

\newcommand{\lastaccessdate}{April 12th, 2022}

\printccsdesc

\section{Trust in Automation}

One of the key challenges of current research and development projects is the automation of the primary driving task. The technology for automated driving is known to be an essential instrument to achieve the goals related to the improvement of road safety~\cite{choi2015investigating}. But what about user acceptance? A low level of technology acceptance is likely to cause an equivalent level of usage and may thus negatively impact the level of market penetration. Beyond technology development, the goal of current research projects therefore is to better understand user acceptance and relevant influencing factors to derive measures that have the potential to contribute to an increase in acceptance. Current studies suggest that potential users both, in Germany~\cite{YouGov2021} and internationally~\cite{vitale20202020} are critical towards highly automated driving. A representative survey conducted by TÜV e.V.~\cite{pols2018mobility} on the assessment and acceptance of autonomous vehicles revealed that 60\% of the respondents foresee increasing safety through use of self-driving vehicles. Nevertheless, 89\% expressed reservations about the new technology. Almost 70\% fear technical problems that are likely to lead to malfunctions or even accidents. 32\% of the respondents trust technology to handle difficult situations less sovereign than a human driver. There is evidence that the listed concerns of the respondents regarding technology acceptance are related to the influencing factor trust in automation (TiA)~\cite{parasuraman1997}.

\begin{figure}[h]%
\centering\subfloat[Gaze tracking as part of a driver monitoring system]{%
\label{fig:gaze_cc}%
\includegraphics[width=0.45\textwidth]{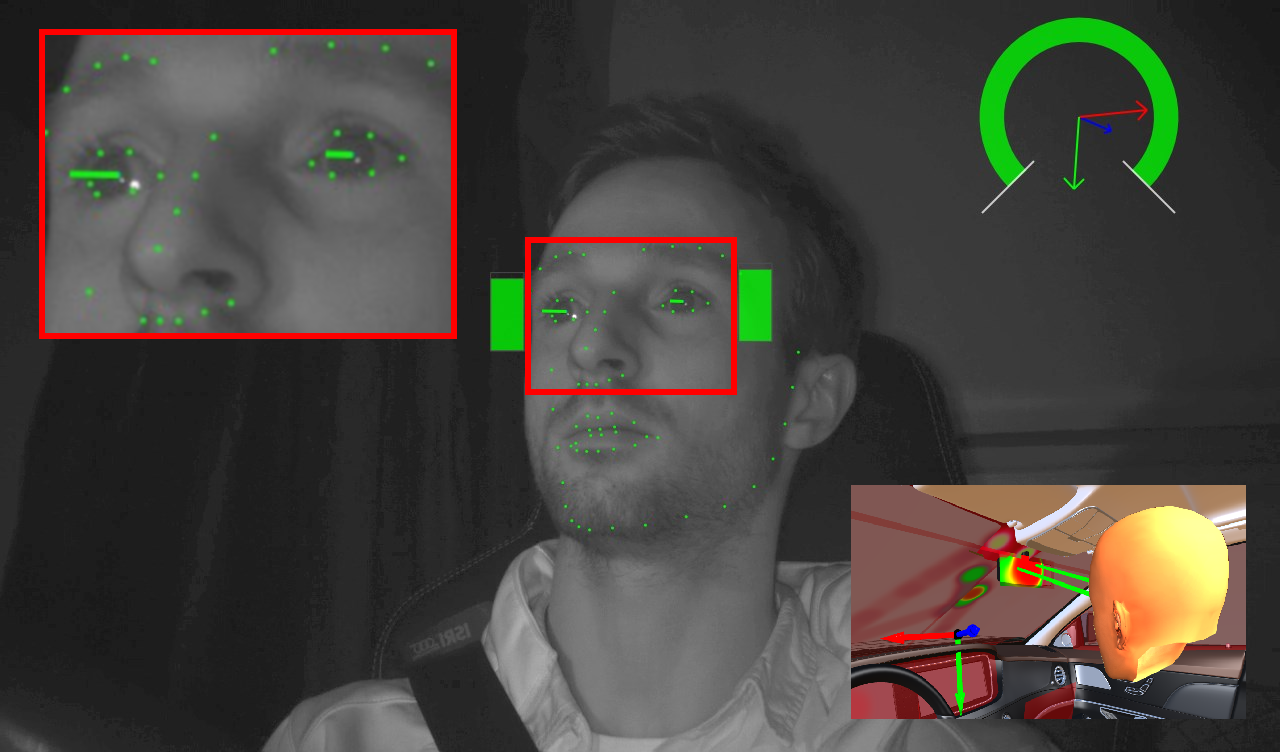}}%
\qquad\subfloat[Emotion expression recognition as key elements for trust recognition]{%
\label{fig:emotion_cc}%
\includegraphics[width=0.45\textwidth]{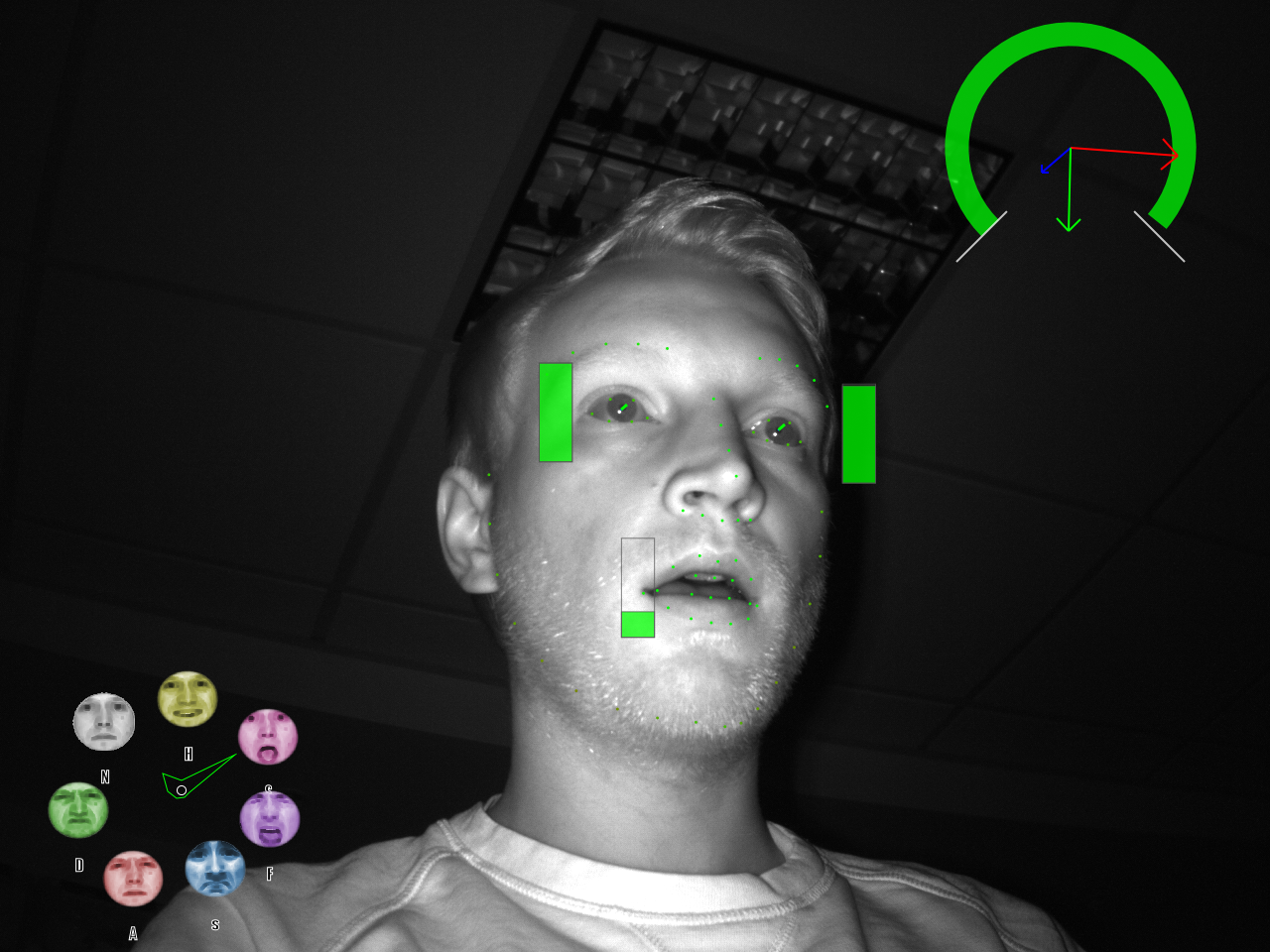}}%
\caption{Development of objective detection of user TiA}
\label{fig:objective_cc}
\end{figure}

 A definition of Lee and See~\cite{Lee2004Trust} describes TiA as \emph{``[…] the attitude that an agent will help achieve an individual’s goals in a situation characterized by uncertainty and vulnerability''}. We likewise understand TiA to be a subjective construct of users that an automated system will act in a desirable manner in situations characterized by uncertainty for and/or vulnerability of the user. TiA not only influences overall acceptance of automated vehicles, but also whether the technology is (in)properly used or even disused~\cite{koerber2018introduction}. An appropriate level of trust therefore seems to be crucial both, in terms of system acceptance and safe system use~\cite{koerber2018introduction}. Due to the relevance of TiA in the context of automated driving, the EMMI project focuses on this construct in particular. First, possibilities for objective detection of users' TiA are aimed at, to recognize the users' socio-emotional behavior during automated driving. An estimation of the users' TiA is discussed to be the central prerequisite for enabling an empathic reaction of the vehicle, which is triggered by fluctuations in the user trust level. The empathic reaction could positively increase the credibility and thus the users' trust in the system in the long term~\cite{scheutz2014artificial}. Overall, multimodal emotion recognition and the associated empathic system response are promising to bear a high unexplored potential to have a lasting impact on trust in automated systems~\cite{Jaimes2007} (see Fig.~\ref{fig:objective_cc}).

Summarizing, the project is pursuing three distinct subsystems that pursue specific goals and approaches to regulate users’ TiA to an appropriate level. The different subsystems comprise an interactive empathic agent, a situational visualization of vehicle status and environment information as well as an indirect vehicle guidance for automated vehicles. The approach and relevance of these goals are described in more detail in the following chapter.
\section{Project Goals}

The three subsystems of the EMMI project follow individual research approaches, all of which can be expected to have an effect on the users' TiA. By integrating them into a comprehensive system, the potentials of the subsystems are combined, and hypothetical synergy effects can arise. The individual research approach and the connected project goals are briefly presented below.

\subsection{Goal 1: Increase trust through an empathic interactive agent that is aware of users' socio-emotional behavior}
\begin{figure}[h]
  \centering
	\includegraphics[width=0.45\textwidth]{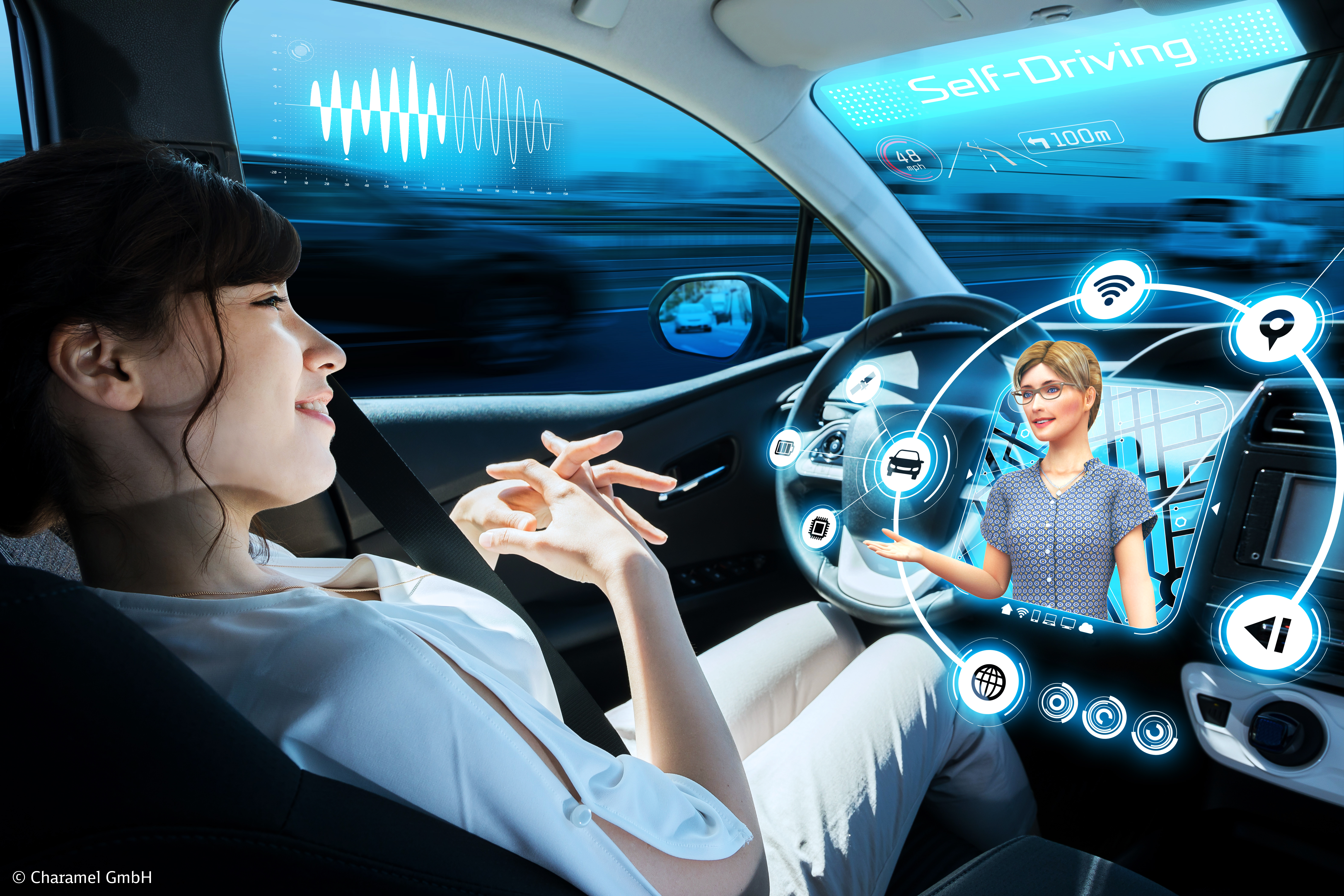}
	\caption{Vision: An empathic interactive agent that supports the user to trust the vehicle automation}
	\label{fig:avatar}
\end{figure}

Since their recent success in smart home technologies, voice-based assistants are now also state of the art in automotive technology. According to forecasts, by 2028 they will be installed in around 90\% of newly registered vehicles worldwide~\footnote{\url{https://www.automotiveworld.com/special-reports/special-report-the-rise-of-the-in-car-digital-assistant/} - last access \lastaccessdate}. The goal of this technology is to realize a natural speech input and output in order to imitate human communication. Adding humanoid aspects and providing the interface with an embodiment can enhance this natural interaction~\cite{pelachaud_studies_2009}. Empathic interactive agents enable human users to interact through communication channels that are natural to them~\cite{cassell_embodied_2000}. Several studies indicate that interactive agents can elicit similar emotions like humans (e.g.~\cite{schneeberger_2019}) and increase trust in automated intelligent systems~\cite{weitz_you_2019}. Also, the use of embodied virtual agents to increase trust in automated driving has been considered~\cite{hauslschmid2017supportingtrust}. Therefore, one goal of the project is to examine the integration of an empathic interactive agent in the context of highly automated driving for user trust management (see Fig.~\ref{fig:avatar}).  

\begin{figure*}[h]
  \centering
	\includegraphics[width=0.85\textwidth]{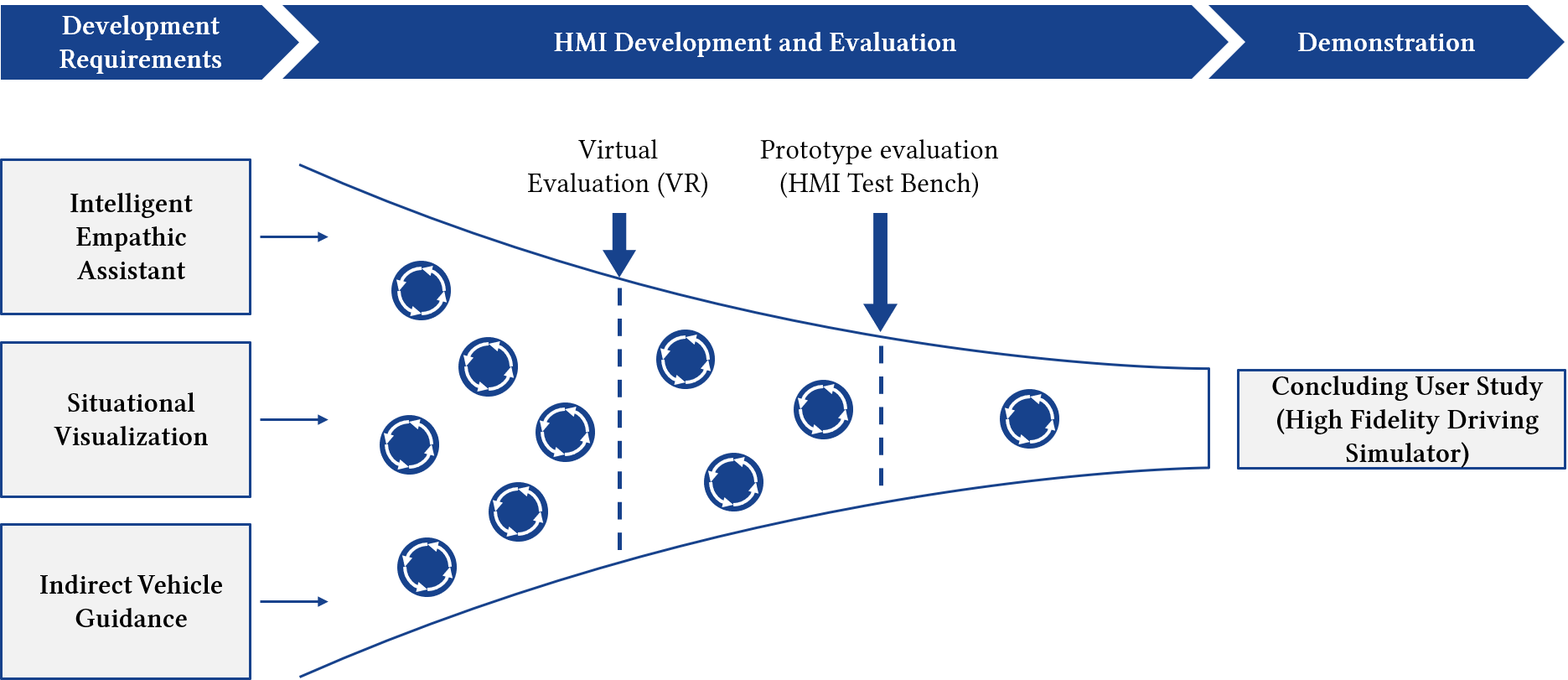}
	\caption{The EMMI Roadmap: Starting with different concepts of three subsystems, these are condensed with use of the multi-stage evaluation process and synthesized into a comprehensive system}
	\label{fig:roadmap}
\end{figure*}

\subsection{Goal 2: Convey understanding for the vehicle behavior with situational visualization of vehicle state and environmental information}
\begin{figure}[h]%
\centering\subfloat[Example: AR visualization of advanced navigation]{%
\label{fig:windshield_sg}%
\includegraphics[width=0.45\textwidth]{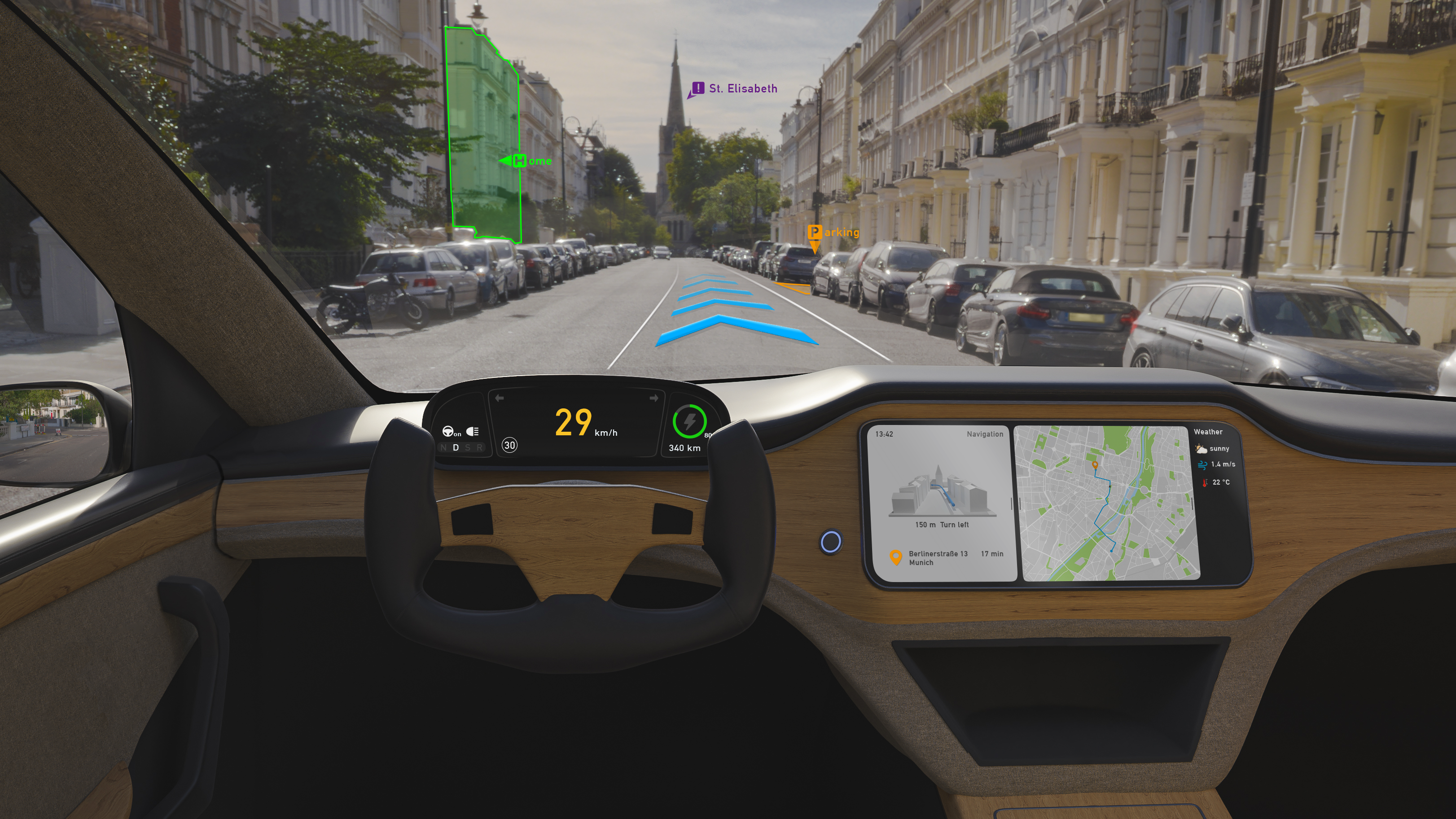}}%
\qquad\subfloat[Example: Concept drawing of trust calibrating visualizations]{%
\label{fig:windshield_ika}%
\includegraphics[width=0.45\textwidth]{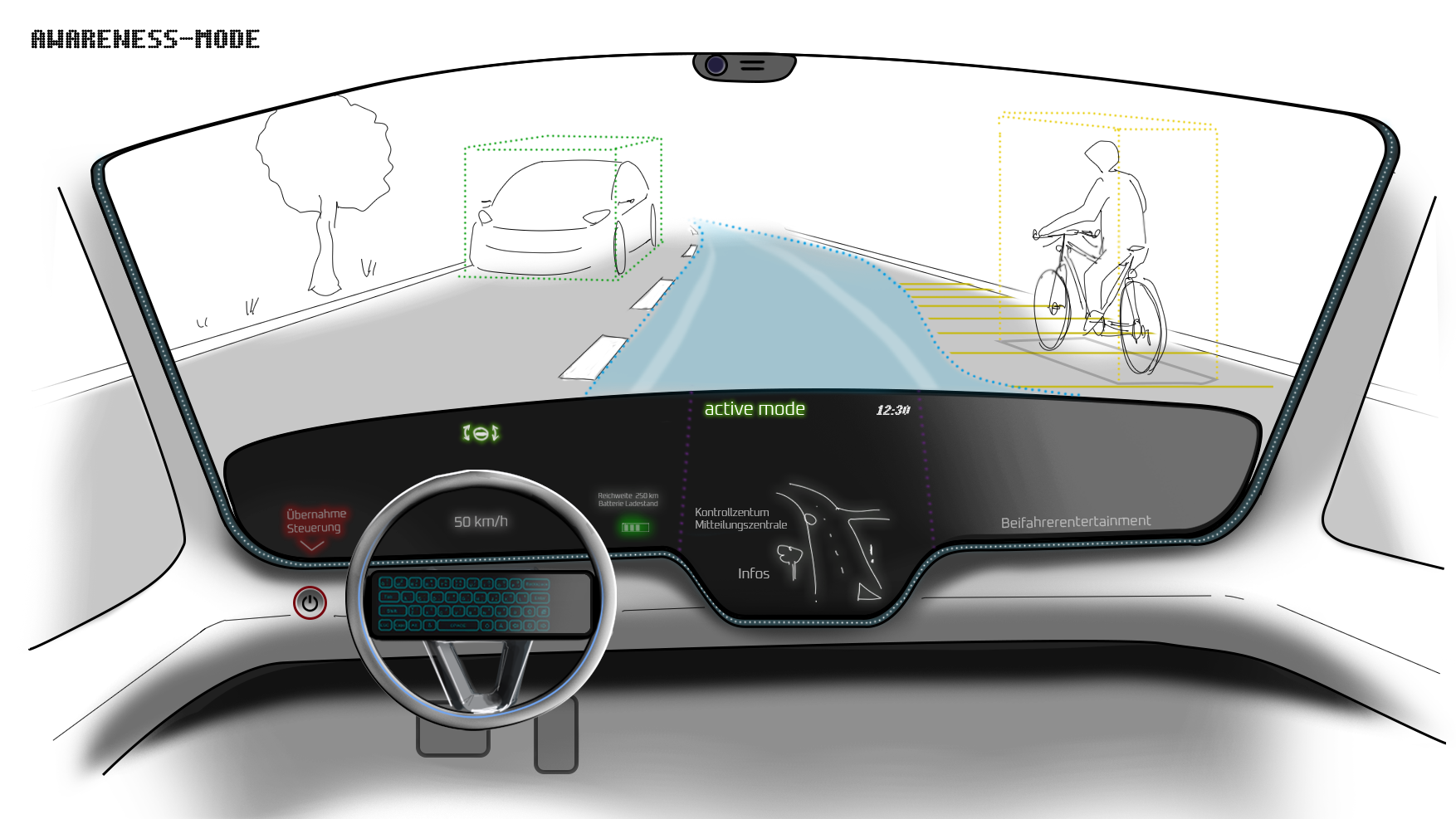}}%
\caption{Vision: Convey understanding of the vehicle automation by presenting adaptive information}
\label{fig:windshield}
\end{figure}

The number of automated research and concept vehicles has been steadily increasing in recent years~\cite{Othman2021}. The increased involvement of novice users raises the question what information the automation has to display and respond to from the users' perspective. The majority of systems currently in use answer this question similarly. To increase the understanding about the automation function, environmental factors, travel trajectories, and detected obstacles in the real environment are specifically presented (see Fig.~\ref{fig:windshield})~\footnote{\url{https://design.google/library/trusting-driverless-cars/} - last access \lastaccessdate}\footnote{\url{https://www.tesla.com/support/full-self-driving-computer} - last access \lastaccessdate}. The underlying research approach is to build trust and acceptance through understanding~\cite{Chang2019,Morra2019BuildingTI}. However, some questions do not seem to have been finally answered yet: What is the main information that can be used to influence the users' level of trust? Is the information able to generate an equal benefit for a broad user population? To what extent should individual user specifics be taken into account with regard to the information to be displayed? If these, and the research questions beyond, remain insufficiently answered with regard to the visualization concepts of highly automated vehicles, in the worst case too much or wrong information might lead to a lack of understanding and thus to uncertainty that might lead to a loss of trust~\cite{Warm1996VIGILANCEAW,Koo2014}. Thus, a systematic investigation of the type, the density and the timing of information, as well as the user group to be considered and its interindividual needs is required to enable a targeted use of display concepts to manage user trust. 

\subsection{Goal 3: Provide the feeling of control by enabling an indirect vehicle guidance for highly automated vehicles}
Various studies have shown that users of automated systems generally trust these systems more if they are given the opportunity to overrule the automation~\cite{Kaur2018,Miller2016}. This increased level of trust is detached from the users' perceived need to override automation. Even in cases where users do not override the automation at any time, a higher basic trust can be found due to the provided possibility of overriding~\cite{Schaefer2016}. Further research has shown that users of automated systems are generally influenced in their manual driving skills~\cite{Bainbridge1983,Seppelt2016}. Due to the lack of situation awareness or calibration of the user as a controller in the driver-vehicle control loop, overriding the automation by the user mostly poses a cognitive challenge and thus a potential threat to road safety~\cite{Endsley1995}. Nevertheless, to realize an appropriate level of trust in automation and to give the user the possibility to influence the automation, an indirect vehicle guidance is needed. The goal is to realize trust by taking into account individual user wishes with regard to automated vehicle control and to return the feeling of control over the vehicle. In addition, an innovative and well-designed interaction can also contribute strongly to the joy of use.

\section{The EMMI Roadmap}
The presented subsystems will subsequently be combined into a comprehensive system and finally evaluated for their effectiveness and UX with respect to trust calibration in the context of highly automated driving. The final investigation will take place in a highly dynamic driving simulator to obtain robust conclusions by providing the greatest possible immersion. The tool makes it possible to examine scenarios that are subjectively perceived as critical without the risk of user intervention that could endanger security. 

\begin{figure}[h]
  \centering
	\includegraphics[width=0.45\textwidth]{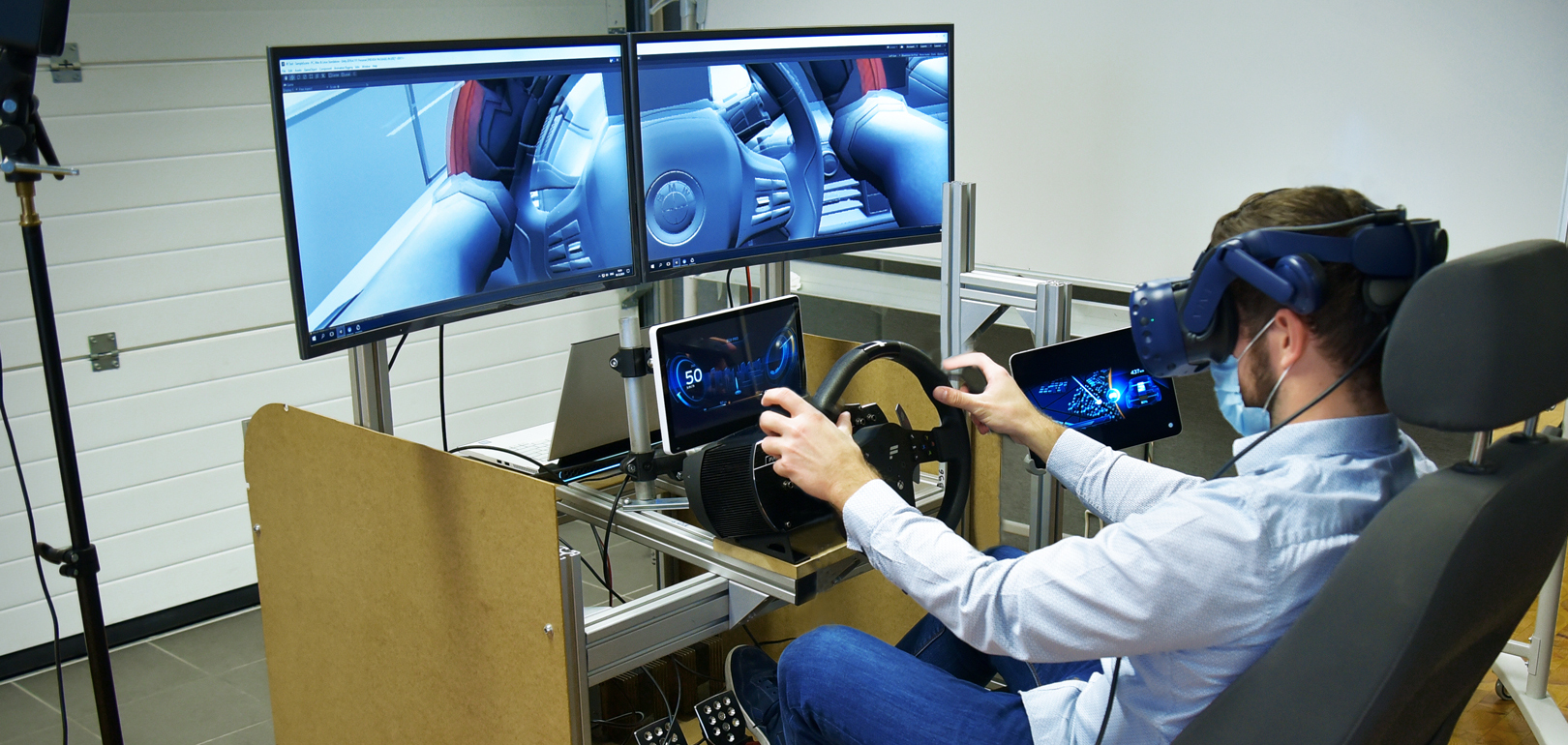}
	\caption{VR environment for early evaluation by the later potential user}
	\label{fig:hmi_lab}
\end{figure}

The project plan envisions that relevant insights into the effectiveness of the individual subsystems will be gained based on the iterative and user-centered development approach even before the final investigation. The evaluation process of the project follows a three stages approach (Fig.~\ref{fig:roadmap}): Initial insights will be gained within a virtual development environment. Thus, the developed concepts will be evaluated in VR close to the later potential user (see Fig.~\ref{fig:hmi_lab}). The insights gained will be iteratively incorporated into the development process. The second stage is the prototype evaluation with functional samples, which will be tested on a new developed HMI test bench. Final findings and approaches for industrialization will be obtained in the concluding user study. 

\section{Acknowledgments}

This work is funded by the German Ministry for Economic Affairs and Energy
in the project EMMI, grant no. 19A20008F.

\balance{}

\bibliographystyle{SIGCHI-Reference-Format}
\bibliography{EMMI_Overview_References}

\end{document}